\documentclass[usegraphicx,usenatbib]{mn2e}

\author[K. Kuijken, X. Siemens \& T. Vachaspati]
{Konrad Kuijken$^1$, Xavier Siemens$^2$, Tanmay Vachaspati$^3$\\
$^1$Leiden Observatory, Leiden University,
PO Box 9513, 2300RA Leiden, The Netherlands\\
$^2$
LIGO Laboratory, California Institute of Technology, Pasadena, 
California 91125, USA;\\
Theoretical Astrophysics, California Institute of Technology, Pasadena, 
California 91125, USA\\
$^3$CERCA, Department of Physics,
Case Western Reserve University, Cleveland, OH~~44106-7079, USA
}

\title{Microlensing by Cosmic Strings}

\begin{document}
\maketitle

\begin{abstract}
We consider the signature and detectability of gravitational
microlensing of distant quasars by cosmic strings. Because of the simple image
configuration such events will have a characteristic light curve, in
which a source would appear to brighten by exactly a factor of two,
before reverting to its original apparent brightness. We calculate the
optical depth and event rate, and conclude that current predictions
and limits on the total length of strings on the sky imply
optical depths of $\la 10^{-8}$ and event rates of fewer than one
event per $10^9$ sources per year. Disregarding those predictions but
replacing them with limits on the density of cosmic strings from the
CMB fluctuation spectrum, leaves only a small region of parameter
space (in which the sky contains about $3\times10^5$ strings with
deficit angle of order 0.3 milli-arcseconds) for which a microlensing
survey of exposure $10^7$ source-years, spanning a 20--40-year period,
might reveal the presence of cosmic strings.
\end{abstract}

\section{Introduction}

Cosmic strings are topological defects that arise during phase
transitions in the early universe, and are also predicted in some
models of inflation (\citealt{kibble76, vilenkinshellard2000,
polchinski2004}). They are dynamical entities, that move at close to
the speed of light. During their evolution strings can intersect each
other (and themselves), which can lead to the formation of closed
loops of strings that subsequently decay. A natural outcome of the
evolution of a primordial network of strings in an expanding universe
is therefore a few long, nearly straight strings that cross the
horizon volume, accompanied by a population of decaying string loops.

The effect of cosmic strings on observational probes of the early
universe has most recently been described by \cite{polchinski2004}.
The most direct way of detecting cosmic strings in the sky is through
their gravitational lensing effect: a string induces a deficit angle
in space, which can lead to double images from sources projected
behind a string. The lensing properties of strings were recently
discussed by \cite{macketal2007} and by \cite{shlaerwyman2005}.

In this letter we consider {\em microlensing by cosmic strings}.
Microlensing (\citealt{liebes1964, refsdal1964, paczynski1986}) occurs
when the source and lens move with respect to one another, leading to
a time-variable image configuration and, consequently, a change in the
total observed flux from the source. It provides a way to detect
lensing even when the image splitting is too small to resolve with
astronomical measurements. Microlensing has developed into a powerful
probe of stellar populations, planetary companions to distant stars,
and dark halos consisting of condensed objects (MACHOs). Particularly
since cosmological constraints have pushed the string tension down to
levels where the deficit angle can be at most a few tenths of
milli-arcsec (\citealt{fraisse2007, wymanetal2005}), microlensing
might offer a feasible alternative route to detecting lensing from
fast-moving strings. Thus the question we address is: can
microlensing by cosmic strings be observed and used as a way of
measuring the string density?

This paper is organized as follows.  In \S2 we describe the image
configuration produced by a cosmic string, and the characteristic
brightness variation expected from string microlensing.  In \S3 we
calculate the expected optical depth, and specific event rate, for a
typical cosmological string population microlensing distant
quasars. \S4 contains a short discussion of the results.

Recently, \cite{chernofftye2007} analysed the expected
  microlensing event rates of stars due to cosmic string loops, and
  concluded that an intensive monitoring campaign would be a sensitive
  probe of an interesting range of string tension. Stellar sources
  offer the great advantage that they have a very compact angular
  extent, but the disadvantage is that the fluxes of individual stars
  can only be monitored to relatively small distances, limiting such
  searches to loops in the halo of the Galaxy, within a few tens of
  kpc from the sun. Unfortunately, a significant population of loops
  in the halo can only arise if the loops manage to lose most of their
  kinetic energy, and this process is not yet properly quantified.

\section{Imaging by cosmic strings}

\begin{figure}
\includegraphics[width=\hsize]{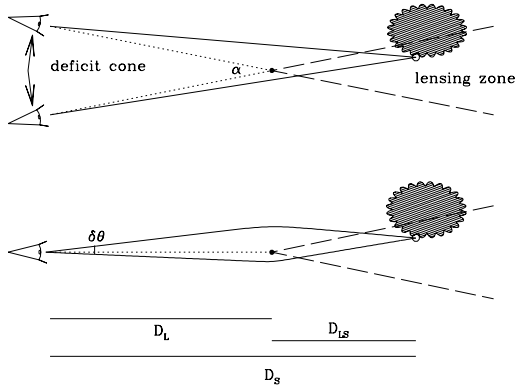}
\caption{Lensing by a cosmic string that passes in front of an
  extended source. In the top diagram we have placed the deficit cone
  induced by the string along the observer-string line, so the dotted
  lines are to be identified. Light rays reaching the observer's eye
  are straight lines in this diagram. In the bottom diagram the same
  configuration is shown, but now the deficit cone has been closed by
  stretching the azimuthal coordinate centered on the string. The
  lensing effect of the string is evident: rays passing either side of
  the string get deflected by an angle $\alpha/2$ towards it.  The
  space between the dashed lines is the lensing zone. The observer
  sees two images of any point in this region, separated by an angle
  $\delta\theta$. On the sky, two copies of the lensing zone are
  visible, separated by the cosmic string.}
\label{fig:bigsrc}
\end{figure}

The effect of a straight cosmic string is to induce a 'deficit angle'
$\alpha$ in space, which has the effect of turning parallel rays that
pass on either side of a string into a converging beam. Effectively
light rays passing the string are bent towards the string by an angle
$\alpha/2$. As a result, background sources that project sufficiently
close to a cosmic string are doubly imaged (see
Fig.~\ref{fig:bigsrc}). 


Because gravitational lensing preserves surface brightness, and
because straight strings do not distort the images they produce, a
doubly imaged source will be exactly twice as bright as the unlensed
source. If a cosmic string were to pass in front of a source
therefore, then the lightcurve would show a sharp step up by a factor
of two, followed some time later by a decrease in brightness back to
the original level. The sharpness of the step is determined by the
angular size of the source: for a point source the step is
instantaneous, but the second image of an extended source will only
gradually be built up (down) as the string tracks across the sky,
smoothing the increase (decline) in brightness.

\begin{figure}
\includegraphics[width=\hsize]{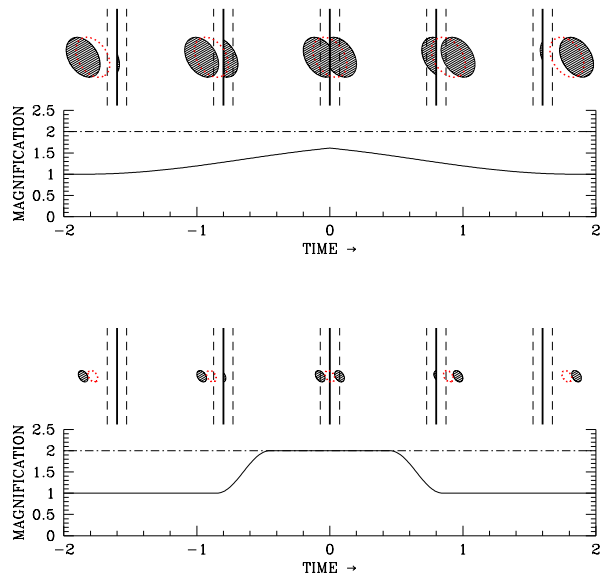}
\caption{
  Image configuration and lightcurve for a source larger than
  the lensing angle $\delta\theta$ (top) and smaller than
  $\delta\theta$ (bottom). The dotted outlines show the
  unshifted locations of the object in our picture of the conical
  space shown in Fig.~\ref{fig:bigsrc}.
  The extra images fall between the dashed
  lines, $\delta\theta/2$ from the string (heavy line). Only for the
  small source will the observed flux reach a plateau of factor-two
  magnification. The large source is never doubly imaged all at once.}
\label{fig:images}
\end{figure}

The key point of microlensing is that this lightcurve can be measured
even without resolving the double images: provided the lightcurve is
sufficiently different from other, astrophysical, variability these
events can be found and studied from large monitoring campaigns of
many objects.

The microlensing described above will only be seen if all of the
source is doubly imaged at once, which means that the source must be
smaller in angular extent than the lensing angle $\delta\theta$.  The
most compact sources that can be seen to cosmological distances are
quasars, whose nuclear regions are $\la1$ parsec in radius,
corresponding to an angular diameter of 0.08
milli-arcsecond\footnote{1 milli-arcsecond is about
$5\times10^{-9}$ radian} (mas) at redshift 1. If
$\delta\theta\simeq\alpha/2$ is smaller, not all of the source can be
doubly imaged simultaneously, and therefore we will not see the full
factor two brightening. Such lightcurves would not show the
characteristic step in brightness either, but a single bump whose
shape and peak amplification is determined by the core structure and
size. We therefore impose the requirement that the flat top of the
light curve lasts at least as long as the rise time (angular diameter
of the source less than $\delta\theta/2$), which implies that we are
sensitive to cosmic strings with deficit angles with
\[
\alpha\ga 0.3 \hbox{mas}
\]
Image configurations and lightcurves for different source sizes are
shown in Fig.~\ref{fig:images}.

As long as the cosmic string is straight on scales comparable to the
source size (and smaller), then the above still holds. Otherwise bends
in the string will spoil the exact factor-of-two magnification, and
may also lead to complex brightness fluctuations as the string tracks
across the source. Such complex lightcurves will be difficult to
recognize as anything other than intrinsic source variability, and so
will not be considered further here.

\section{Optical depth and event rate}

\subsection{Event duration}

A string with deficit angle $\alpha$ (Fig.~\ref{fig:bigsrc}) that lenses
a point source will produce two images that are separated by an angle
\[\delta\theta=\alpha (D_{LS}/D_S)\]
where $D_{LS}$ is the angular diameter distance from source to string,
and $D_S$ is distance to the source. Such image splitting takes place
if the source lies within a projected angle $\delta\theta/2$ of the
string, so if the string moves in front of the source, at transverse
speed $V$, the duration $T$ of the double-image phase is given by
\[
T=\delta\theta D_L /V = {\alpha D_L D_{LS}\over V D_S}. 
\]
Typically (for all reasonable cosmological models including the
  current standard model of a flat universe with
  $\Omega_\Lambda\simeq0.7$), $D_{LS}/D_S$ will be about 0.5 (for a string
at redshift $\sim$1, and a source at redshift $\ga2$), so that the
image splitting is about half of the deficit angle. The distance $D_L$
to a string at redshift 1 is about $c/H_0$ where
$H_0\simeq(14\times10^9\hbox{yr})^{-1}$ is the Hubble constant. A
typical value for $V$ is around 0.3 times the speed of light
\citep{vilenkinshellard2000}.  Thus the duration $T$ of a microlensing
event, from beginning of ingress to the beginning of egress, is of the
order of
\[
T \sim {\alpha\over 0.6 H_0} \simeq 120 \left(\alpha\over 1\hbox{mas}\right)
\hbox{yr}.
\]
\label{sec:est-T}
A similar calculation shows that the time over which the source
brightness doubles (or halves again) is about 20 years, for a 0.08mas
source size.

Very similar lensing behaviour would result from dense gas
  filaments in the intergalactic medium, provided their angular width
  on the sky is thinner than the source size and than their own
  deficit angle of $8\pi G\lambda/c^2$, for linear mass density
  $\lambda$ \citep{Bozza2005}. Indeed, the same applied to filaments
  of dark matter. However, because such filaments would move a factor
  100-1000 times more slowly than cosmic strings, the microlensing
  timescales involved would be 100--1000 times longer.

\subsection{Optical depth}

The {\em optical depth} $\tau$ for lensing is the probability that a
given source is doubly imaged at any given time, and is therefore
equal to the fraction of sky that is covered by strips of angular
width $\delta\theta$ along all visible strings.

A typical long string crossing our horizon volume will have a
projected length on the sky of $\sim\pi$ radians.  If there are $N$
such cosmic string in the sky, then the area of sky that is
multiply-imaged is about $\delta\theta \times \pi N$ radians, which
represents a fraction $N\delta\theta/4$ of the sky. Hence
\[
\tau={N\alpha D_{LS}\over4D_S}\simeq 
6\times10^{-10}N \left(\alpha\over 1\hbox{mas}\right)
\]
where the estimate is based on the same assumptions as in
\S\ref{sec:est-T}. 

This tiny probability means that a huge number of background
sources, sufficiently compact (angular size $\ll\delta\theta$) to be
lensed significantly, would have to be observed in order to have a
reasonable chance of detecting one lensed source.

\subsection{Event rate}

The {\em event rate} $\Gamma$ at which a given source will enter the
microlensing regime is given by the lensing probability divided by the
typical duration:
\[
\Gamma=\tau/T={NV\over4D_L}\simeq {N\over4}{V\over c}H_0
\simeq 5\times10^{-12} N \hbox{yr}^{-1}.
\]
Note that the lensing rate is independent of the deficit angle
$\alpha$: it is purely determined by the rate at which cosmic strings
sweep regions of sky, which is a function only of their angular
diameter distance, transverse velocity, and total length. As the
transverse velocity is of the order of the speed of light, the angular
motions on the sky for strings at cosmological distances are of the
order of the Hubble constant.

\section{Discussion}
The calculation above shows that the event rate to be expected from a
network of straight cosmic strings is tiny: a typical source at high
redshift will be microlensed by a cosmic string of order $N$ times per
Hubble time, where $N$ is the number of long strings that exist in our
horizon volume. Predictions derived from simulations of the simplest
variety of strings (\citealt{hindmarshkibble1995,
vilenkinshellard2000}) put $N$ at 10--100, leading to a very low event
rate indeed: $10^9$--$10^{10}$ sources would have to be monitored for
a year in order to see a few events.

Furthermore, the duration of the events is of the order of 40 years,
even for the lightest strings (those with deficit angles $\sim0.3$mas)
that are heavy enough to microlens parsec-sized cores of distant
quasars, so it would require a very sustained campaign to observe a
full lightcurve. Typical values of the rise (or decay) time are 20
years. Only if a suitable background population of bright, numerous,
compact (sub-pc) sources at redshifts above $\sim2$ can be identified
can strings with smaller deficit angles, with correspondingly shorter
lensing timescale, be probed. The timescale would also be shortened if
there were a population of cosmic strings at low redshift.

There are suggestions that networks of cosmic superstrings---which
have the same lensing properties as `ordinary' cosmic strings---evolve
differently because they are less likely, perhaps by as much as a
factor of 1000, to intercommute when they cross (see \citealt{tye} and
references therein). This may lead to an enhancement by up to a factor
of 1000 in the length of string on the sky, i.e. $N$. Even with this
boost the event rate remains small, at one event per several million
sources per year at best. It remains to be seen whether future radio
surveys will contain a sufficiently large number of compact sources
that can be monitored for microlensing events.

Thus far we have concentrated on long-lived, horizon-crossing strings,
but it is also possible that the length of string in loops is much
larger than that in long strings. Loops with radii much larger than
$\delta\theta$ lens distant sources in the same way as straight
strings.  Whether there is a substantial loop population or not
depends on details of cosmic string network evolution that have yet to
be resolved.  Currently there are two possibilities.  The first is
that loop sizes are determined by gravitational back-reaction. If that
is the case then loops are short-lived and the length of string in
loops is at all times comparable to the length in long strings
\citep{vilenkinshellard2000}.

The second possibility, suggested by recent numerical simulations, is
that loop sizes are determined by the large-scale dynamics of the
network (\citealt{Ringeval, Martins, VOV}).  In this case loops may be
long-lived, slowly shrinking in length $L$ by emission of
gravitational radiation at a rate $dL/dt\sim-\alpha c$
\citep{vilenkinshellard2000}.  Assuming that the largest loops to form
have sizes $\sim10\%$ of the horizon size $ct_f$ \citep{VOV}, for a
formation time $t_f$ they will survive to the present time $t_0$
provided $\alpha\la 0.1 t_f/t_0$. Thus, because the loop density in
the radiation era is much larger than in the matter era, large numbers
of string loops with $\alpha\la 10^{-7}$ formed around or before the
end of the radiation era could still exist.  It turns out they do not
enhance the microlensing rate significantly: using the loop length
distribution from Eq.~(70) in \citet{SCMMCR} we find that the total
length in loops scales as $\alpha^{-1/2}$, and even in the most
favourable case of $\alpha\sim10^{-9}$ it only reaches the equivalent
of $N\simeq50$.

Perhaps these predictions for $N$ will turn out to be unduly
pessimistic, so let us assume that somehow a very large length of
cosmic string has managed to survive to redshift
$\sim1$. Our calculations of the event rate have shown that in this
case a cosmic string microlensing survey of several million sources
might find some events. But in this case we run into another
observational limit: the strong constraint that arises from the
spectrum of fluctuations in the cosmic microwave background (CMB). As
shown by \cite{wymanetal2005}, the density fluctuations arising from
strings ($\delta_s$) must be less than $\sim 10\%$ of the primordial
adiabatic density fluctuations $\delta_a \sim 10^{-5}$.  The density
fluctuations due to strings are characterized by rms fluctuations in
the number of strings. Therefore, as the total energy density per
horizon-crossing string is $\sim\alpha$ times the critical density,
$\delta_s \sim \sqrt{N}\alpha < 0.1 \times 10^{-5}$.  As discussed
above, for sources of physical radius $1 {\rm pc}$ a Hubble distance
away, $\alpha \ga 1.5\times10^{-9}$ is required for the characteristic
microlensing lightcurve to be produced. So the maximum possible number
of strings is $N_{\rm max} \sim 3\times10^5$, and the maximum
observable event rate is $\Gamma_{\rm max} \sim1.5\times 10^{-6} {\rm
yr}^{-1}$.  Even in this most favourable case, millions of sources
would need to be monitored to observe a few events per year. 

\begin{figure}
\includegraphics[width=\hsize]{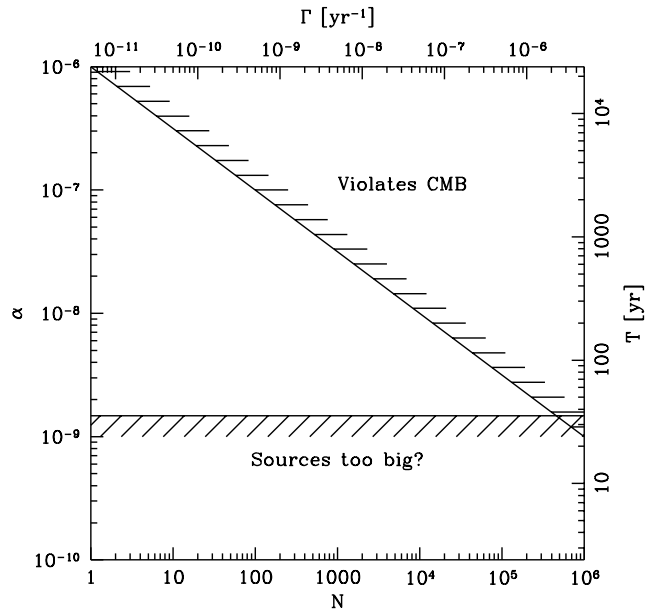}
\caption{Available parameter space (deficit angle $\alpha$ in radians,
  vs.\ string length $N$) for microlensing surveys. The top axis gives
  the event rate expected, the right-hand axis the event duration. For
  parameters above the diagonal line the strings would generate
  fluctuations in the cosmic background radiation inconsistent with
  observations. Below the horizontal line sources as small as 1pc in
  radius (angular diamater 0.08mas at redshift 1) would be too large
  in angular extent to be completely lensed. For details see the
  text.}
\label{fig:limits}
\end{figure}

These limits, which are independent of the cosmic string formation
scenario, are plotted in Fig.~\ref{fig:limits}.
We conclude that the prospects for detecting cosmic strings through
simple microlensing of quasars are bleak!

\section*{Acknowledgments}

We thank Ana Ach\'ucarro, Tom Kibble, Irit Maor, Huub R\"ottgering,
Ignas Snellen, and Daniel Wesley for discussions. T.V.\ thanks the
Netherlands Organization for Scientific Research (NWO) for a visitor's
grant, and the Lorentz Institute of Leiden University and the ICTP in
Trieste for hospitality; his work was supported by the U.S.\
Department of Energy and NASA at Case Western Reserve University. The
work of X.S. was supported in part by NSF Grant PHY-0601459. LIGO was
constructed by the California Institute of Technology and
Massachusetts Institute of Technology with funding from the National
Science Foundation and operates under cooperative agreement
PHY-0107417.

\end{document}